\documentclass[aps, prb, reprint, twocolumn, floatfix]{revtex4-1}

\usepackage{graphicx}
\usepackage{epstopdf}
\usepackage{gensymb}

\begin{document}

\title{A dispersive nanoSQUID magnetometer for ultra-low noise, high bandwidth flux detection}
\author{E. M.  Levenson-Falk}
\author{R. Vijay}
\author{N. Antler}
\author{I. Siddiqi}
\email[Corresponding author: ]{irfan@berkeley.edu}
\affiliation{Quantum Nanoelectronics Laboratory, Department of Physics, University of California, Berkeley CA 94720}

\date{\today}

\begin{abstract}

We describe a dispersive nanoSQUID magnetometer comprised of two variable thickness aluminum weak-link Josephson junctions shunted in parallel with an on-chip capacitor.  This arrangement forms a nonlinear oscillator with a tunable 4-8 GHz resonant frequency with a quality factor Q = 30 when coupled directly to a 50 $\Omega$ transmission line. In the presence of a near-resonant microwave carrier signal, a low frequency flux input generates sidebands that are readily detected using microwave reflectometry. If the carrier excitation is sufficiently strong then the magnetometer also exhibits parametric gain, resulting in a minimum effective flux noise of 30 n$\Phi_0$/Hz$^{1/2}$ with 20 MHz of instantaneous bandwidth. If the magnetometer is followed with a near quantum-noise-limited Josephson parametric amplifier, we can increase the bandwidth to 60 MHz without compromising sensitivity. This combination of high sensitivity and wide bandwidth with no on-chip dissipation makes this device ideal for local sensing of spin dynamics, both classical and quantum.       
\end{abstract}


\maketitle

Superconducting quantum interference devices (SQUIDs) constructed from nanoscale weak-links\textemdash nanoSQUIDs\textemdash are routinely used in magnetization studies of single magnetic molecules and ferromagnetic clusters \cite{Wernsdorfer2009, Russo2011,Lam2003,Hao2011,Troeman2007}. The weak-link constriction geometry permits efficient electromagnetic coupling to nanoscale magnets\cite{nano_coupling_vincent}, thus suggesting the possibility of localized single-spin detection with a superconducting electrical circuit. Currently, nanoSQUIDs consist of unshunted planar structures operated in the vicinity of the voltage state. As such, these devices effectively operate as a magnetic-flux-dependent switching current detector with measured flux noise $S_\Phi^{1/2}$ (defined as the smallest signal resolvable per unit bandwidth) of order $\mu\Phi_{0}/$Hz$^{1/2}$ and kHz of bandwidth. The latter is limited by the reset time of the unshunted SQUID to the zero-voltage state at subkelvin temperatures. Moreover, the dissipation and junction dynamics associated with switching to the voltage state can produce significant measurement backaction\cite{nanosquid_back_moler}, potentially making this mode of operation incompatible with quantum state detection.

We consider instead a dispersive magnetometer that operates as a nonlinear oscillator with a flux- and power-dependent resonant frequency. The device is biased by applying a microwave frequency carrier tone, and read out by comparing the phase of the reflected carrier signal with the input. A low frequency flux signal appears as a phase modulation of the carrier. The weak-link junctions remain in the zero-voltage state, thus eliminating on-chip dissipation and its associated backaction. A previous prototype device based on tunnel junctions achieved an effective flux noise of $S_{\Phi}^{1/2}=$ 290 n$\Phi_0$/Hz$^{1/2}$ with 20 MHz bandwidth\cite{hatridge_PRB}. In our present work, we apply the principle of dispersive readout to an optimized nanoSQUID device and obtain $S_{\Phi}^{1/2}=$ 30 n$\Phi_0$/Hz$^{1/2}$ with 20 MHz bandwidth. These values are observed using a strong carrier pump that results in 20 dB of power gain, sufficient to overcome the noise associated with the first stage semiconductor microwave amplifier. If instead the magnetometer is immediately followed by a near quantum-noise-limited Josephson parametric amplifier \cite{hatridge_PRB, Vijay_jumps}, we can reduce the magnetometer gain without a diminution in sensitivity and boost the bandwidth to 60 MHz. Thus, this dispersive nanoSQUID has flux sensitivity on par with the best reported tunnel junction SQUIDs\cite{VanHarlingen1982}, tens of MHz of instantaneous bandwidth, no on-chip dissipation, and a constriction based geometry; these features make it ideal for localized sensing of nanoscale magnets, including spin based classical logic or quantum bits.

A false-colored scanning electron micrograph of the device is shown in Fig. 1. Our nanoSQUID is comprised of 3D weak-links where the banks are significantly thicker than the bridge. These variable thickness bridges, as compared with 2D (Dayem) planar structures, exhibit enhanced nonlinearity and a current-phase relation approaching that of an ideal point contact \cite{Vijay_dcwire,Eli_rfwire}.  This results in greater transduction of an input magnetic flux to a microwave frequency voltage, as the inductance of the nanoSQUID (and thus the resonant frequency of the tank circuit) changes more effectively with flux.  The bridges are 100 nm long, 30 nm wide and 20 nm thick, and connect banks which are 1.5 $\mu$m long, 750 nm wide and 80 nm thick. The structure is fabricated using electron beam lithography and double angle evaporation of aluminum.  The two-junction nanoSQUID has a critical current $I_C \approx 20 \ \mu$A and is shunted by a 7 pF parallel plate capacitor (with an amorphous silicon nitride dielectric) to form a 4-8 GHz flux-tunable nonlinear resonator \cite{Eli_rfwire}. We couple the resonator directly to a 50 $\Omega$ transmission line to form a low quality factor $Q=30$ circuit. This allows us to bias the resonator in the nonlinear regime for improved flux sensitivity, as detailed later, while maintaining sufficient bandwidth for high speed operation \cite{hatridge_PRB}. Unlike other SQUID based magnetometers, there is negligible dissipation in the nanoSQUID and shunting capacitor, allowing low noise operation without any associated heating or the need for cooling fins \cite{squid_heating_MSA_cooling_fins}. A terminated section of coplanar waveguide transmission line near the nanoSQUID is used to couple a test flux signal ranging in frequency from dc up to several GHz. The chip is mounted on a microwave circuit board with an integrated $180\degree$ hybrid, over a superconducting coil used to apply large static flux. This structure is placed inside superconducting and cryoperm magnetic shields to minimize external magnetic noise, and cooled to 30 mK in a dilution refrigeration.  

\begin{figure}
\includegraphics{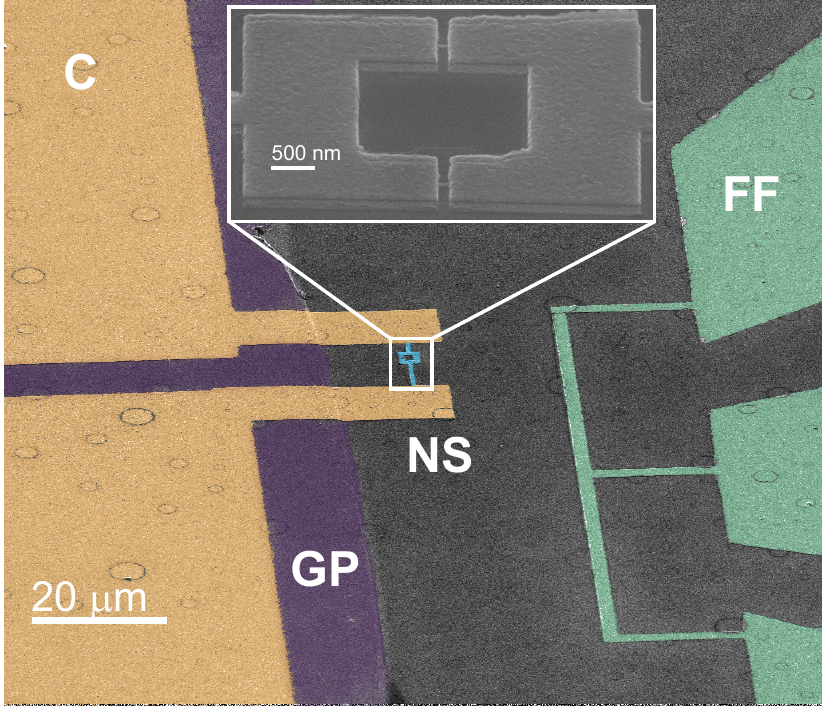}
\caption{\label{fig1} (Color online) False-colored scanning electron micrograph of the dispersive nanoSQUID magnetometer.  The nanoSQUID (NS, light blue, detailed in the inset) is shunted by two parallel-plate capacitors (C, orange), comprised of aluminum top electrodes over a niobium ground plane (GP, purple) with a silicon nitride dielectric in between.  Flux signals from dc up to GHz frequency, used to calibrate sensitivity, are coupled via a short-circuit terminated section of 50$\Omega$ transmission line (FF, light green).}
\end{figure}

\begin{figure}
\includegraphics{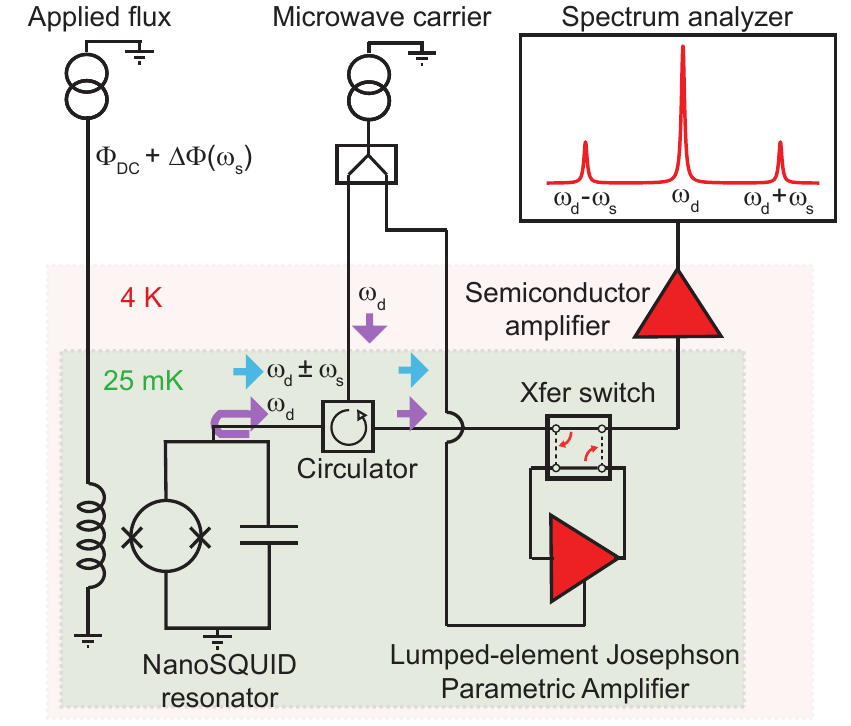} 
\caption{\label{fig2} (Color online) Simplified schematic of the experimental apparatus.  The magnetometer consists of a nanoSQUID shunted by a capacitor, mounted on the 30 mK cold stage of a dilution refrigerator.  A microwave carrier tone at $\omega_d$ reflects off the device through a circulator, then passes through multiple amplification stages before being recorded at room temperature.  A superconducting coil provides static magnetic flux bias, while an on-chip fast flux line couples in low-frequency flux signals at $\omega_s$.  The magnetometer upconverts flux signals to carrier sidebands at $\omega_d\pm\omega_s$, which are also amplified before measurement at room temperature. A parametric amplifier (LJPA), driven by a copy of the same carrier tone, may be switched in or out of the measurement chain \textit{in situ}, allowing for near quantum-noise-limited amplification.  The frequency response is recorded using a spectrum analyzer; the height of the sidebands above the noise floor determines the signal-to-noise ratio and thus the effective flux noise.}
\end{figure}

A simplified measurement setup is shown in Fig. 2. A circulator, which allows signals to propagate in one direction (arrows), is used to reflect microwaves off the magnetometer. The reflected signal is then amplified by a Lumped-element Josephson Parametric Amplifier (LJPA) \cite{hatridge_PRB, Vijay_jumps}. The LJPA serves as a preamplifier with noise performance near the quantum limit, allowing us to reduce the overall system noise and thereby improve the measurement signal-to-noise ratio (SNR).   The signal is further amplified by a cryogenic semiconductor amplifier and several room temperature amplifiers before being sent to a spectrum analyzer for final analysis. The LJPA preamplifier can be bypassed \textit{in situ} using a cryogenic switch to compare the operation of the nanoSQUID under different amplification configurations. The superconducting coil is used to apply a static flux bias to the nanoSQUID while the on-chip fast flux line allows coupling of calibrated, low frequency flux signals to benchmark magnetometer sensitivity.  To operate the device, we first pump it with a microwave tone at $\omega_d$.  This tone reflects off the device and acquires a phase shift that depends on the nanoSQUID resonant frequency.  At finite static flux bias, the resonant frequency is a sensitive function of applied flux, and thus the phase shift provides a measure of the flux threading the nanoSQUID. In the presence of a single frequency ac flux signal at $\omega_s$, the resonant frequency (and thus the phase shift) oscillates at $\omega_s$.  In the frequency domain, this phase oscillation is observed as two perfectly correlated microwave sidebands at $\omega_d \pm \omega_s$. A sample spectrum is shown in Fig. 2.  

The nanoSQUID resonator has two modes of operation: the linear and the parametric regimes. When the carrier excitation is weak ($<$ -75 dBm), the magnetometer is essentially a harmonic oscillator with a flux-dependent resonant frequency. In this linear mode of operation, the magnetometer can be viewed as an upconverting transducer of low frequency flux signals to the microwave domain without any power gain (symbolized in the legend of Fig. 3 as a circular transducer). The peak height of the microwave sidebands is a linear function of the input flux signal $\Delta\Phi$ and the transduction factor $dV / d\Phi $.  The effective flux noise is obtained by comparing the magnitude of the transduced signal with the overall noise level of the microwave amplification chain (including quantum noise). Specifically, for a given flux excitation ($\Delta \Phi$), we measure the voltage signal-to-noise ratio ($V_{\rm{SNR}}$) of one of the upconverted sidebands within a given integration bandwidth ($B$) and calculate the effective flux noise using the relation $S^{1/2}_{\phi} = \Delta\Phi/(V_{\rm{SNR}} \ (2 B)^{1/2})$.  The factor of 2 in the denominator is to account for the fact that both sidebands contain information while the noise is uncorrelated. Fig. 3 shows the effective flux noise as a function of the flux excitation frequency when operated in the linear regime (green circles). Here, the LJPA has been bypassed. We obtain a flux noise of 210 n$\Phi_0$/Hz$^{1/2}$ with about 100 MHz of instantaneous bandwidth. In the linear regime, the bandwidth is limited by the resonator quality factor and is consistent with the measured Q = 30 of our 6 GHz oscillator. The semiconductor amplifier has a bandwidth of 4 GHz and does not limit the operational bandwidth. However, the 8 K system noise temperature of the amplification chain (dominated by the low temperature semiconductor amplifier) limits the effective flux noise, as it introduces excess voltage noise onto the microwave signals.  At low frequencies (below 1 kHz) we observe intrinsic flux noise with a 1/f character and a value of $\sim 1 \ \mu \Phi_0$/Hz$^{1/2}$, as is typical in SQUID sensors\cite{Wellstood1987}.

\begin{figure}
\includegraphics{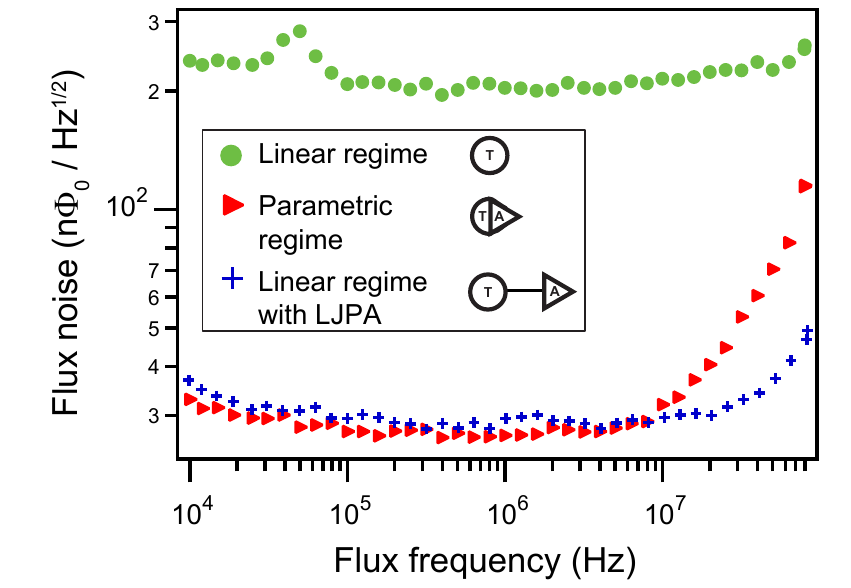} 
\caption{\label{fig3} (Color online) Effective flux noise of a dispersive nanoSQUID in the linear and parametric regimes.  In the linear mode with no following LJPA, the device has a flux noise of 200 n$\Phi_0 / $Hz$^{1/2}$ with a large bandwidth of $\sim$100 MHz.  In the parametric mode, the device gain serves to overcome the noise from the following semiconductor amplifier, but at the cost of reducing bandwidth.  Here, the flux noise is 30 n$\Phi_0 / $Hz$^{1/2}$ and the bandwidth is $\sim$20 MHz. When the LJPA is switched in, the system noise nears the quantum limit, and flux noise reaches the same 30 n$\Phi_0 / $Hz$^{1/2}$ value with a broader bandwidth of $\sim$60 MHz set by the LJPA. }
\end{figure}

In the parametric regime, we utilize the fact that the nanoSQUID is a nonlinear inductor at high excitation energy, resulting in parametric amplification of the upconverted flux signals\cite{Eli_rfwire}. In this mode, we can model the magnetometer as a two-stage device: a transducer followed by a near-noiseless parametric amplifier (paramp; symbolized in Fig. 3 as a combined circular transducer and triangular amplifier)\cite{hatridge_PRB}. This ultra-low noise amplification process lowers the effective system noise temperature of the measurement chain and improves the sensitivity of the magnetometer significantly.  Fig. 3 shows the effective flux noise (red triangles) in the parametric mode, which was calculated using the procedure described in the previous paragraph. Note however that we do not divide by the factor of 2 in this case since both sidebands now have identical information (signal and noise) due to correlations associated with parametric amplification \cite{hatridge_PRB}. We now obtain an effective flux noise of 30 n$\Phi_0$/Hz$^{1/2}$ but at the expense of a reduced bandwidth of 20 MHz. This reduction is characteristic of parametric amplifiers involving a resonant circuit where the product of the voltage gain and instantaneous bandwidth is conserved \cite{hatridge_PRB}. At this level of power gain, the overall system noise temperature is quantum-limited. As such, additional improvements in SNR can only be obtained by increasing the magnitude of the transduced microwave signal, since the noise floor cannot be lowered any further.

The transduction factor $dV / d\Phi$ increases with the carrier drive amplitude $V_{\rm{in}}$ and the flux dependence of the resonance frequency $df_{\rm{res}}/d\Phi$.  Thus, flux noise may be lowered by increasing the microwave carrier strength and tuning the static flux bias farther away from zero.  However, the onset of resonator bifurcation at high drive powers limits the maximum carrier amplitude.  This threshold is further reduced at finite flux bias where the critical current of the nanoSQUID drops and its effective nonlinearity is enhanced\cite{Eli_rfwire}. In practice, we find the lowest-noise operating points in the neighborhood of $\Phi = \Phi_0 / 4$.  It is noteworthy that the critical drive amplitude for bifurcation depends linearly on the junction critical current\cite{vijay_thesis}; thus, the maximum carrier power can be increased using higher critical current junctions.  In fact, we attribute the decrease in flux noise in this weak-link device relative to the previous tunnel-junction prototype\cite{hatridge_PRB} primarily to a factor of 5 increase in the junction critical current.  Further optimization is possible, although increasing the critical current beyond its current value will decrease the Josephson inductance to a value necessitating a redesign of the full circuit to minimize the effects of stray geometric (linear) inductance associated with the on-chip bias lines, SQUID loop, and capacitor pads.

In the parametric mode, there is a complication that reduces flux sensitivity. The resonator operates as a phase sensitive amplifier where one quadrature of an input ac signal is amplified noiselessly, while the other quadrature is de-amplified \cite{Vijay_jumps}. The upconverted microwave voltage created by the transduction stage of the magnetometer represents a single quadrature signal, but its axis is not perfectly aligned with the amplified quadrature of the phase sensitive amplifier \cite{hatridge_PRB}. As a result, only the component parallel to the amplified quadrature gets amplified, resulting in reduced sensitivity. This process is illustrated in Fig. 4(a-b).  The relative angle $\theta_t$ between transduction and amplification axes is an intrinsic property of the nonlinear oscillator (which is well-described by the Duffing model) and approaches a value of 60 degrees for large parametric gain \cite{hatridge_PRB, claflamme1}. This implies that the output signal is a factor of two smaller\textemdash i.e. effective flux noise is a factor of two higher\textemdash than the theoretical limit which can be achieved if one could can the relative angle $\theta_t$. In effect, the transduction has been reduced, since we can only measure the portion of the upconverted signal which lies along the amplified quadrature. Such a degradation does not occur when the device is operated with unity gain (i.e. in the linear regime). However, the high noise temperature of the amplification chain  limits sensitivity in that mode and large paramp gain is essential to ensure that the dominant source of noise in the experiment is quantum fluctuations of the vacuum.  Thus, there is a tradeoff between reduced transduction and reduced noise as the paramp gain rises.

To quantify this effect, we implemented the following experiment. The frequency of the microwave carrier was chosen to provide a maximum power gain G of about 20 dB (voltage gain of 10). We swept the signal power and measured the associated paramp gain by reflecting an additional weak microwave tone, near the carrier frequency, off the biased nonlinear resonator. This voltage gain is plotted in Fig. 4(c) (blue line, right axis) and reaches a maximum value of about 13. At each carrier power, we also inject a 1 MHz flux tone and measure the amplitude of one of the upconverted sidebands. This amplitude is proportional to the magnitude of the flux tone, the transduction coefficient ($dV / d\Phi \ \cos \theta_t$), and the paramp gain. We divide the sideband amplitude by the magnitude of the flux tone and the paramp gain to infer a value we term the $t$-factor, which is equal to the transduction coefficient times the net voltage gain of the measurement chain ($\sim 10^5$ in our setup): $t = dV / d\Phi \ \cos \theta_t \ \sqrt{G_{\rm{sys}}}$. This quantity is plotted in Fig. 4(c) as the solid red line (left axis). At low bias powers, the transduction initially increases but then drops quickly by about a factor of two, as expected, as the gain increases. In the linear regime, the transduction is directly proportional to the bias amplitude but in the paramp regime, the relative angle $\theta_t$ starts to increase with gain. 

\begin{figure}[t]
\includegraphics{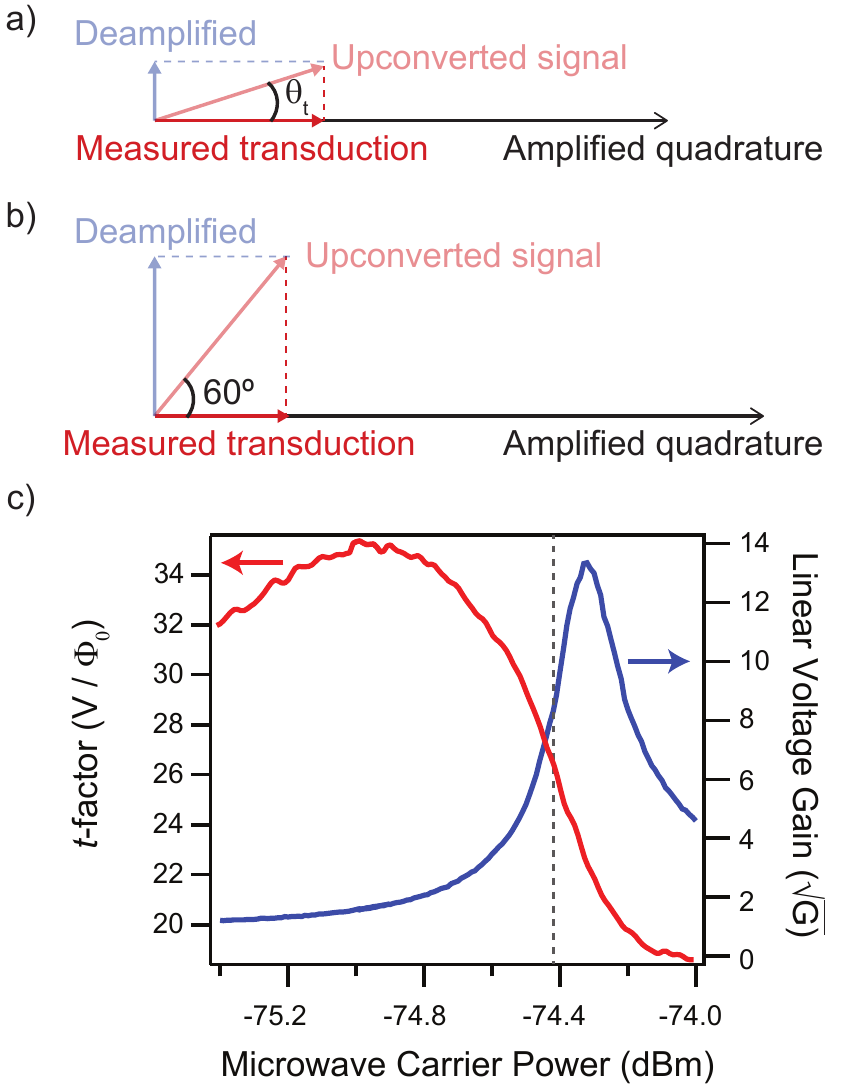} 
\caption{\label{fig4} (Color online) (a-b) Illustration of reduced transduction with increasing gain.   At intermediate gain (a), the phase of the upconverted signal starts to change relative to the drive (center); only the component in phase with the drive is amplified, while the other component (ghosted blue arrow) is de-amplified.  Only the amplified component (red arrow) is measurable, and so it determines the effective transduction. At high gain (b) the upconverted signal grows with power, but is now 60 degrees out of phase with the pump, so the transduction involves only half of the upconverted signal. Thus, transduction falls, even as the upconverted signal grows. (c) Flux transduction $t$-factor (transduction times system gain) and gain of the device as a function of microwave drive power.  The transduction (red, left axis) has a maximum at the highest power where gain (blue, right axis)  is still small\textemdash i.e. where $dV / d\Phi$ is large but $\theta_t$ is small.   The dashed line indicates the power at which flux noise is minimized, optimizing the tradeoff between large transduction and low noise temperature (high gain).}
\end{figure}

In order to optimize transduction and amplification independently, we made use of the LJPA amplifier. In this configuration, we could reduce the gain of the nanoSQUID without being limited by the noise of the subsequent amplification chain. First, we maximized bandwidth by operating the nanoSQUID in the linear regime (defined here as power gain less than 1 dB) and the LJPA with 20 dB gain (symbolized in the legend of Fig. 3 by a circular transducer connected to a triangular amplifier); at this bias point, this particular amplifier had a 60 MHz bandwdith. We obtained an effective flux noise of 30 n$\Phi_0$/Hz$^{1/2}$ with an improved bandwidth of 60 MHz as shown in Fig. 3 (blue crosses). Future improvements in amplifier technology can further increase the instantaneous bandwidth since the intrinsic nanoSQUID bandwidth in the linear regime can readily reach $\sim$ 100 - 200 MHz.  To optimize the system for flux noise, we next operated the nanoSQUID at the point of maximum transduction. We then adjusted the LJPA such that the combined gain of both stages was 20 dB and adjusted $\theta_t$ for optimum amplification.  This resulted in an effective flux noise of 23 n$\Phi_0$/Hz$^{1/2}$, a noticeable improvement over single-stage operation. We could not, however, obtain the factor of two improvement in flux noise that was predicted.  We believe this is due to transmission losses between the magnetometer and the LJPA, as the transduced signal propagates from the first stage to the second via a series of passive components. These transmission losses can be as high as 2 - 3 dB depending on the number of circulators and other microwave components used \cite{fb_rabi}, thus nullifying some of the SNR improvement. By minimizing this loss, it should be possible to approach optimal flux noise performance. We also observed a reduction in bandwidth from 60 to 20 MHz. As shown in Fig. 4(c), in this device, optimal transduction actually occurs at a drive power where there is finite parametric gain and thus a reduction of bandwidth. In future devices, the strength of the nonlinearity can be diluted by increasing the critical current or shunting with a linear inductor to suppress this effect.  

In conclusion, we have demonstrated an ulta-low noise dispersive magnetometer with the potential to detect single electron spins with high bandwidth and negligible dissipation. Further work is needed to characterize the resilience of this device to in-plane magnetic fields \cite{Antler_2013} often necessary to obtain a desired energy level splitting in nanoscale magnets.  Finally, we note that the two stage system we have realized provides a controlled test bed to characterize measurement induced backaction on a weakly coupled quantum system, in this case a magnetic spin, due to a nonlinear detector. Such an experiment could be realized by measuring the coherence time of a coupled spin as a function of the magnetometer gain, thus providing a direct route to explore recent theoretical predictions \cite{Clerk2010,claflamme1}.  

Financial support was provided by AFOSR under Grant FA9550-08-1-0104, the NSF E3S center under NSF award EECS-0939514, and DARPA under Grant N66001-09-1-2112. We also acknowledge support from the NSF GRFP (EMLF and NA). 

\bibliography{nanoSQUIDmag}

\end{document}